\documentclass{aastex631}
\usepackage{hyperref}
\usepackage{xcolor,multirow,array,booktabs,amsmath,amssymb,natbib,mathrsfs}
\usepackage{graphicx,epstopdf,appendix,threeparttable,chngpage,txfonts,esint}
\usepackage{ulem}

\newcommand{\fexiii}{Fe \textsc{xiii}}
\newcommand{\heii}{He \textsc{ii}}
\usepackage{soul}

\begin{document}

\title{Early Evolution of the Cavity and Core of a Coronal Mass Ejection in the Inner Corona}

\author[0000-0003-2694-2875]{Shuting Li}
\affiliation{Key Laboratory of Dark Matter and Space Astronomy, Purple Mountain Observatory, \\
Chinese Academy of Sciences, Nanjing 210023, People's Republic of China}
\affiliation{School of Astronomy and Space Science, University of Science and Technology of China, \\
Hefei 230026, People's Republic of China}

\correspondingauthor{Li Feng}
\author[0000-0003-4655-6939]{Li Feng}
\email{lfeng@pmo.ac.cn}
\affiliation{Key Laboratory of Dark Matter and Space Astronomy, Purple Mountain Observatory, \\
Chinese Academy of Sciences, Nanjing 210023, People's Republic of China}
\affiliation{School of Astronomy and Space Science, University of Science and Technology of China, \\
Hefei 230026, People's Republic of China}

\author[0000-0001-8402-9748]{Beili Ying}
\affiliation{Key Laboratory of Dark Matter and Space Astronomy, Purple Mountain Observatory, \\
Chinese Academy of Sciences, Nanjing 210023, People's Republic of China}

\author[0000-0001-5705-661X]{Hongqiang Song}
\affiliation{School of Space Science and Technology, Shandong University, \\
Weihai, Shandong 264209, People’s Republic of China}
\affiliation{State Key Laboratory of Space Weather, National Space Science Center, \\
Chinese Academy of Sciences, Beijing 100190, \\
People’s Republic of China}

\author[0000-0001-7397-455X]{Guanglu Shi}
\affiliation{Key Laboratory of Dark Matter and Space Astronomy, Purple Mountain Observatory, \\
Chinese Academy of Sciences, Nanjing 210023, People's Republic of China}
\affiliation{School of Astronomy and Space Science, University of Science and Technology of China, \\
Hefei 230026, People's Republic of China}

\author[0009-0001-4778-5162]{Jiahui Shan}
\affiliation{Key Laboratory of Dark Matter and Space Astronomy, Purple Mountain Observatory, \\
Chinese Academy of Sciences, Nanjing 210023, People's Republic of China}
\affiliation{School of Astronomy and Space Science, University of Science and Technology of China, \\
Hefei 230026, People's Republic of China}

\author[0000-0003-3160-4379]{Jie Zhao}
\affiliation{Key Laboratory of Dark Matter and Space Astronomy, Purple Mountain Observatory, \\
Chinese Academy of Sciences, Nanjing 210023, People's Republic of China}

\author[0000-0001-9979-4178]{Weiqun Gan}
\affiliation{Key Laboratory of Dark Matter and Space Astronomy, Purple Mountain Observatory, \\
Chinese Academy of Sciences, Nanjing 210023, People's Republic of China}
\affiliation{University of Chinese Academy of Sciences, Nanjing, China}

\begin{abstract}

Coronal mass ejections (CMEs) typically exhibit a three-component structure in white-light (WL) coronagraphs. Utilizing the seamless observations of the inner corona ($\le$ 3~R$_\odot$), we have revealed the early evolution of the cavity and core of a CME starting at $\sim$18:20~UT on 2014 October 14. The CME originates from a hot channel (HC), which appears as the bright core and compresses the cavity in WL images. Specifically, most of the dark cavity in WL is filled by bright loop-like structures in 174~\AA. The differential emission measure (DEM) analysis indicates that the electron temperature decreases from the core ($\sim$13.4~MK) to the cavity ($\sim$1.35~MK), and the CME cavity is significantly cooler than that enshrouding a prominence ($\ge$ 2~MK). The effective temperature of the cavity increases over time in general, probably due to the compression by the HC expansion. The evolution of the CME bright core includes slow-rise, fast-rise (up to $\sim$330~km s$^{-1}$), and residual-acceleration phases. The cavity exhibits an evolution similar to the core but lags by $\sim$4 minutes, with a lower speed peaking at $\sim$220~km s$^{-1}$. Moreover, the 2D radial speed distribution exhibits the highest speeds at the core apex. The kinematical results further confirm the compression of the cavity. The present event supports the new explanation of the CME structures, i.e., the magnetic flux rope (MFR), which is proxied by the HC, is only responsible for the core, while the cavity is likely a low-density region between the CME front and the MFR.

\end{abstract}


\section{Introduction} \label{sec:intro}

CMEs are large-scale eruptions on the Sun. They can carry large amounts of plasma and magnetic field into the interplanetary space and cause geomagnetic storms when they arrive at the Earth and interact with the magnetosphere \citep{Gosling_1991}. Their related space weather may affect the high-technology systems around the Earth. Observations of the CMEs with WL coronagraphs demonstrate that 30\% of CMEs appear as a three-part structure, i.e., a bright front, a dark cavity, and a bright core \citep{Illing_observation_1985,Webb_1987SoPh,Webb_2012_LRSP}, corresponding to a high-low-high density sequence. \citet{Song_2019b} suggested that the three-part configuration of CMEs is more frequently observed during the early evolution -- all four CMEs they investigated possess three components in the low corona at EUV wavelengths, regardless of whether a distinct cavity is observed when the CMEs propagate into the coronagraph field of view (FOV).

It is widely believed that CMEs originate from the eruption of MFRs that form either prior to \citep{Patsourakos_2013, Kliem_2021ApJ, Yao_2024ApJ} or during \citep{Cheng_2011, Song_2014} the eruptions. The MFR is a coherent magnetic structure with a bundle of field lines wrapping around the central axis and showing twist \citep{TD_1999A&A,Chen_2011LRSP}. \citet{Zhang_2012NatCo} reported an MFR observed as a HC in AIA 131~\AA\ and 94~\AA\ that existed approximately 10~minutes prior to the CME eruption for the first time. Their study also identified a bright front in AIA 171~\AA, resulting from the compression by the expansion of the MFR. However, the counterpart of the cool compression front as seen by the WL coronagraph is not well understood due to the limited FOV of AIA. \citet{Cheng_2013ApJ} suggested that the HC is most likely the MFR and acts as a continuous driver of the CME eruption during the early acceleration phase. To date, it is widely accepted that the HCs, which are found to be hot and dense structures, are the proxy of the MFRs in the inner corona \citep{Cheng_2014ApJa,Song_2015ApJ}.

In the traditional opinion, the three components of the CME respectively correspond to the prominence/filament, the MFR, and the plasma pileup along the boundary of the expanding MFR \citep{Vourlidas_2013SoPh}. In other words, the entire cavity is regarded as the MFR; the denser filament at the dip of the MFR forms the bright core. However, the nature of CME structures has yet to reach a consensus. Observations of some CMEs unrelated to filaments challenged the traditional opinion \citep{Howard_2017ApJ,Song_2017}. \citet{Song_2019a} confirmed that both the HC and filament material can appear as the bright CME core in the WL coronagraph. To this end, \citet{Song_2017,Song_2022ApJ,Song_2023ApJL} proposed a new scenario to interpret the nature of the cavity and core. They suggested that the MFR only corresponds to the bright core, while the low-density zone between the front and MFR -- rather than the expanding MFR itself -- is responsible for the dark cavity. Recently, \citet{Song_2025ApJ} provided more observational evidence supporting the new explanation.

The kinematic evolution of CMEs usually consists of three phases: a slow-rise phase with approximately constant velocity, a main-acceleration phase lasting from several minutes up to several hours, and a post-acceleration phase with only a weak residual acceleration \citep{Zhang_2004,Maricic_2004,Maricic_2009}. The first two phases often occur in the inner corona. Moreover, the three-phase evolution is found to be coupled with the associated flares, i.e., the pre-flare phase, impulsive phase, and decay phase in soft X-ray (SXR) flux \citep{Neupert_eruption_2001,Cheng_2010}. The slow magnetic reconnection before the onset of flares might be responsible for the storage of magnetic free energy, thus the slow rise of the CME. The torus instability or flare reconnection drives the main acceleration \citep{Cheng_2013ApJ,Cheng_2020,Lin_2000JGR,Green_2018SSRv}. The main acceleration phase is followed by the simple propagation phase which has either a nearly constant velocity or a residual acceleration \citep{Cheng_2010ApJ}. In general, the correlation between the kinematic evolution of CMEs and the SXR flux curve of flares indicates the same energy releasing process is going on, transferring the magnetic energy into kinematic or thermal energy. 

The thermodynamic properties can imply the heating or cooling mechanisms that occurred in the CME during the evolution. The capability of multi-temperature and high temporal resolution of AIA promises the ability to diagnose the multi-thermal properties of CMEs in the low corona. The DEM can be applied to reconstruct 2D distribution of the electron density and temperature. Many studies have proposed and developed the DEM codes to derive the DEM solution with solely AIA EUV images \citep{Hannah_2012A&A,Weber_2004IAUS,Cheung_2015,Su_2018,LiZhentong_2022}. \citet{Cheng_2012} applied the DEM method to study the multiple structural components of three CMEs, they found that the core (presumably the magnetic flux rope of the CME) has the highest average temperature (8.29-10.89~MK) and density ($\sim$10$^9$~cm$^{-3}$), the relatively cool temperature of $\sim$2~MK appears at bright leading front (LF) in EUV wavelengths. They concluded that the core regions are heated by magnetic reconnection, and the LFs result from the compression of ambient corona plasma as the MFR expands. The highest temperature in the CME core is also found by \citet{Bemporad_2007} with the UV spectra acquired by the Ultraviolet Coronagraph Spectrometer onboard \textit{Solar and Heliospheric Observatory}. The other interesting result in their work is the higher temperature in the leading edge with respect to the ambient corona. The plasma heating provided by magnetic dissipation and adiabatic compression are taken as the main explanation for the high temperature in the core and the leading edge, respectively. Nevertheless, the thermal properties of CMEs can vary considerably due to their different nature. The temperature of the core can range from 0.8-2.5 MK for a CME associated with filament, to 8.0-15.8~MK for that associated with MFR but lacking filament \citep{Chmielewska_2016AN}.

The prominence cavities observed off the limb have been investigated in a few important works. Spectral line imaging observations covering a broad temperature range of 0.9~MK to 2~MK during total solar eclipses, together with EUV imaging and coronagraph WL observations, provide unique opportunities to reveal the thermal and magnetic structures of prominence cavities \citep{Habbal_2010ApJ, Habbal_2014ApJ}. \cite{Habbal_2010ApJ} discovered from eclipse observations for the first time that the prominence cavities have intricate twisted magnetic structures, with a temperature of 2~MK or even higher for the most part of the cavity enshrouding low-lying prominences, while with a relatively low temperature for the cavity enshrouding suspended higher prominences. The higher temperature that ranges from 1.67~MK to 2.15~MK was found in prominence cavities in the AIA FOV with the DEM method as well \citep{Baksteslicka_2019SoPh}. The hot material in the suspended prominence cavities would cool down to 1~MK due to expansion \citep{Habbal_2014ApJ}, which is consistent with the electron temperature in the expanding corona reported by \cite{Habbal_2010ApJL}. The nature of the intricate magnetic structure of the prominence cavity has been further confirmed in \cite{Habbal_2014ApJ} by combining eclipse observations together with EUV images. The analysis in their study proved that the complex turbulent-like structures filling with approximately 2~MK plasmas are dominated within the immediate vicinity of the prominence, forming the cavities. Moreover, a number of models have predicted the helical magnetic structures of the prominence cavities \citep{Fan_2006ApJl,Fan_2019FrASS,Schmit_2014IAUS}. On the other hand, the WL polarized brightness and EUV observations reveal that the intensity depression in cavities is only relative to the immediate streamer environment but is comparable to the intensity of the background corona \citep{Fuller_2008ApJ,Fuller_2009ApJ}. Moreover, the appearance of the helical twisted structures depends on their orientation along the line of sight (LOS) \citep{Shaik_2024ApJ}. 

However, the kinematic and thermodynamic properties of cavities surrounding the HC are less investigated, especially during the early evolution of CMEs. Though the CME event occurred on 2014 October 14 has already been studied by some previous works \citep{Witasse_2017JGRA,West_2015ApJ,Song_2023ApJL}, the targets of these studies are either the related interplanetary coronal mass ejection \citep{Witasse_2017JGRA} or the post-eruptive loops \citep{West_2015ApJ}. \citet{Song_2023ApJL} presented this event as one sample of their statistical work to demonstrate that more CMEs should possess three-component configuration in the low corona. Nevertheless, the detailed evolution properties of this event have not been studied yet. All the above considerations motivated the work presented here, we provide in-depth understanding of the core and cavity of a CME by taking advantage of the seamless observations of the CME from the solar disk up to 3~R$_\odot$ in different passbands, e.g., the EUV wavelengths, the WL, and near infrared. For the first time, we report the observational evidence of the heating inside the cavity caused by the compression due to the expansion of the HC, as well as the direct evidence supporting the new explanation of the nature of the cavity. The description of instruments and the related methods are given in Section~\ref{sec:Methods}. Section~\ref{sec:Results} introduces the observations and results, which is followed by the summary and discussion (Section~\ref{sec:Summary}).

\section{instruments and methods} \label{sec:Methods}
\subsection{Instruments}
The Atmospheric Imaging Assembly \citep[AIA; ][]{Lemen_2012} instrument on board the Solar Dynamics Observatory \citep[SDO; ][]{Pesnell_2012} images the Sun in seven EUV wavelengths with a FOV up to 1.3 $R_{\odot}$. The AIA has a spatial resolution of about 0.6$^{\prime\prime}$ per pixel and a cadence of 12 seconds. In this work, we primarily use the 131~\AA\ to display the HC, and perform DEM calculations with the remaining five wavelength bands (94~\AA, 171~\AA, 193~\AA, 211~\AA, and 335~\AA), excluding the 304~\AA. The Sun Watcher using Active Pixel System detector and Image Processing \citep[SWAP; ][]{Seaton_2013SoPh,Halain_2013SoPh} instrument on board the PRoject for Onboard Autonomy 2 \citep[PROBA2;][]{Santandrea_2013SoPh} provides images of the solar disk and lower corona at about 174~\AA, with a cadence of 1–2 minutes and a FOV up to 1.6~$R_\odot$. The K-coronagraph \citep[K-Cor; ][]{dewijin_2012} is one of instruments in the COronal Solar Magnetism Observatory \citep[COSMO; ][]{Tomczyk_2016JGRA} facility suite. The K-Cor records the polarization brightness (pB) of lower corona with a FOV from 1.05 to 3 $R_\odot$ at a cadence of 15 seconds. We use K-Cor data to understand the formation, the structure, and the evolution of the CME in the low corona. The Coronal Multichannel Polarimeter \citep[CoMP; ][]{tomczyk_instrument_2008} instrument located at the Mauna Loa Solar Observatory has a FOV from about 1.03 to 1.5 $R_\odot$ with a spatial resolution of 4.5$^{\prime\prime}$ per pixel. The CoMP records the intensity as well as the linear and circular polarization (Stokes I, Q, U, V) of the forbidden lines at \fexiii\ 1074.7/1079.8~nm. In addition to detecting the field direction in the plane of sky (POS) and the field strength along the LOS, CoMP also measures the dynamic properties of the plasma from the Doppler and line width observations. We mainly use the CoMP data to derive the Doppler velocity along LOS and the effective temperature of CME structures.

\subsection{Methods}
\subsubsection{DEM method}
DEM analysis, performed using AIA EUV images, is a critical diagnostic tool for understanding the thermal properties of various structural components of CMEs. Since the \heii\ line in AIA 304~\AA\ channel is often optically thick, while DEM analysis relies on optically thin emissions. So that this line is not well-modeled and unsuitable for DEM analysis. We use the six near-simultaneous AIA EUV images (94~\AA, 131~\AA, 171~\AA, 193~\AA, 211~\AA, 335 \AA) to calculate the DEM. The observed count rate ($y_i$) of each passband can be determined by:
\begin{equation}\label{eq:dem}
     y_i=\int K_i(T) \times DEM(T)\mathrm{dT}
\end{equation}
where $K_i(T)$ represents the temperature response function of the $i$-th passband, $DEM(T)$ is the plasma DEM as a function of temperature. With the DEM analysis, we can derive two quantities, i.e., the DEM-weighted average temperature $\overline{T}$ and the total emission measure (EM). $\overline{T}$ is defined as follows:
\begin{equation}
    \overline{T} = \frac{\int_{T} DEM(T)T\mathrm{dT}}{EM}
\end{equation}
where $EM= \int_{T} DEM(T) \mathrm{dT}$, indicating the column density of the plasma along the LOS. We use the xrt\_dem\_iterative2.pro routine \citep{Weber_2004IAUS} in the Solar Software (SSW) library to perform the DEM calculation. On the other hand, \citet{saqri_2020} found that some instrument effects, e.g., the point spread function \citep[PSF; ][]{Grigis_PSF_2012} of AIA/EUV data can affect the DEM temperature diagnosis, especially for the relatively faint structure in higher temperature. To guarantee a better accuracy and reduce the error of DEM result, the AIA data we used are deconvoluted with a PSF model. The uncertainties of the DEM results are estimated with 100 Monte Carlo (MC) solutions. For each MC solution, the input observed intensities are modified by adding a random normal Gaussian distribution with 1~$\sigma$ equals the uncertainty in the observed intensity. Then, the DEM is solved again using the new set of intensity. So that the 100 MC solutions represent the estimated uncertainties of the best-fit DEM solution, indicating how well the DEM is determined.

\subsubsection{Estimate of the effective temperature}
We estimate the approximated effective temperature from the \fexiii\ line width observations acquired by CoMP. Considering that the line broadening occurring in the corona is mostly the combination of the Doppler broadening and the non-thermal broadening \citep{Del_2018}. \cite{Tu_1998} suggested the following formula to calculate the upper limit of the effective temperature $T_{i,\rm{eff}}$ of ions of species $i$:
\begin{equation}\label{eq:Teff}
    V_{i,\rm{eff}}^2 = \frac{2k_B T_{i,\rm{eff}}}{m_i}=\frac{2k_B T_i}{m_i}+\xi^2
\end{equation}
where $k_B$ is the Boltzmann constant, $T_i$ is the temperature of the ions, $m_i$ is the mass of the ion \fexiii, $\xi$ represents the turbulence velocity accounting for the non-thermal broadening. $V_{i,\rm{eff}} = \frac{c}{\lambda} \Delta\lambda_i$ is the effective ion speed derived from the CoMP observation of the Doppler half-width $\Delta\lambda_i$, where $c$ is the speed of light, $\lambda$ represents the rest wavelength of the spectral line. Considering the relationship between the line width and the thermal or non-thermal speed of ions, it is common to express the line width in units of speed. The broadening induced by instrumental resolution limits and noise, i.e., the instrument width, should be removed from the measured line width to provide the thermal and non-thermal width. In our calculations, the instrument width of CoMP (21~km s$^{-1}$) as estimated by \citet{Morton_2015NatCo} has been subtracted.

\subsubsection{Kinematic analyses}
The early kinematics of the CME in the inner corona is understood by taking a slice along the eruption direction. Stack plots are constructed by combining time-series slices extracted along the CME nose using AIA, SWAP, and K-Cor observations. The height-time profiles of CME structures therefore can be measured visually upon the stack plots. The height-time profiles are usually fitted by different functions that might link to different acceleration mechanisms. \cite{Cheng_2020} found that the slow-rise phase and the subsequent main-acceleration phase can be best fitted by a linear and an exponential function, respectively, corresponding to the tether-cutting and torus instability. To this end, we also perform the fitting for the height-time profiles of the CME core and cavity. The speed-time profiles are derived by calculating the first derivative of the height-time measurements.

Furthermore, the Cross-Correlation Method \citep[CCM; ][]{Ying_2019} based on the WL observation can be applied to derive the 2D radial projected velocity distribution in the POS. The CCM requires three consecutive sequences ($T_{n-1}$, $T_n$, $T_{n+1}$) of WL images in polar coordinates (the horizontal axis is the radial distance and the vertical axis is the latitude). It consists of three steps, including a forward step (FS, by using images at $T_n$ and $T_{n+1}$), a backward step (BS, by using images at $T_n$ and $T_{n-1}$), and an average step (AS, averaged by the speed obtained from the FS and BS). At a fixed latitude, we can acquire the center-pixel displacements of the signal window at $T_n$ pixel by pixel based on the maximal cross-correlation between $T_n$ and $T_{n+1}$ (or $T_{n-1}$) through the FS (or BS). Subsequently, the displacement can be converted to the radial speed at $T_n$ for each pixel divided by the time interval between the two consecutive frames. Finally, the speeds derived from the FS and BS are averaged in the AS.

\section{Observations, analyses, and results}\label{sec:Results}
\subsection{Morphology} \label{subsec:morphology}
\begin{figure}[ht!]
    \centering
    \includegraphics[width=1.0\linewidth]{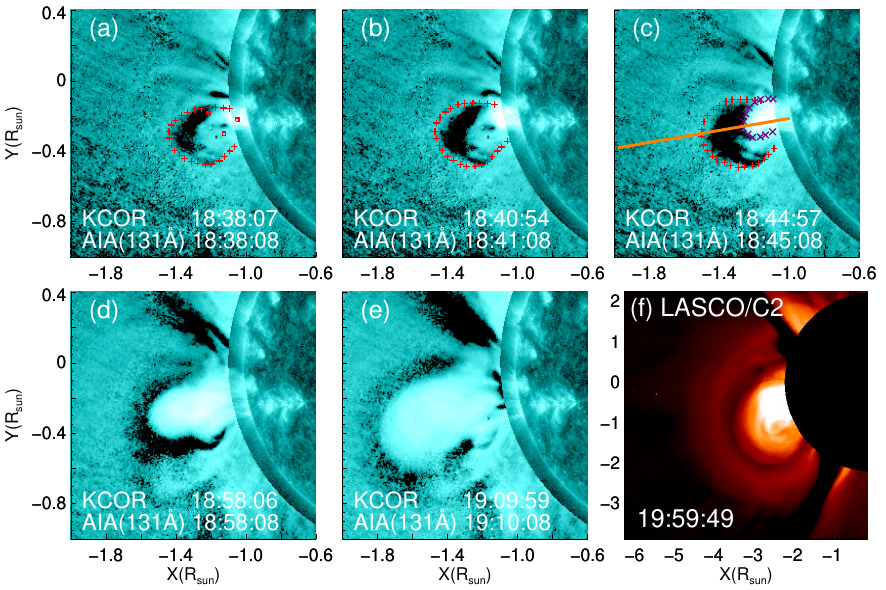}
    \caption{(a)-(e) Base-difference K-Cor images at different times during the eruption. The red boxes in panel (a) delineate the selected sub-regions for the calculation of DEM curves. The red pluses in panels (a)-(c) represent the outer edge of the loop structures in SWAP 174~\AA\  (Figures~\ref{fig:SWAP} (b)-(d)). The purple pluses in (c) outlines the edge of HC observed in AIA 131~\AA, while the orange solid line denotes the position of a slice at which the stack plots (Figures~\ref{fig:slice}) are constructed. (f) The observation of the CME in LASCO/C2, showing the CME configuration in the high corona. An animation of base-difference images from K-Cor, AIA 131~\AA, and 193~\AA\ (from left to right in the animation) is available, showing the evolution of the CME from 18:00~UT to 19:14~UT on 2014 October 14.}
    \label{fig:WLimages}
\end{figure}

A CME is observed by K-Cor in the inner corona from 1.05~R$_\odot$ to 3~R$_\odot$ and the Large Angle and Spectrometric Coronagraph \citep[LASCO; ][]{Brueckner_1995SoPh} C2 in the outer corona from 2.2~R$_\odot$ to 6~R$_\odot$ on 2014 October 14. The coordinated data analysis workshops (CDAW) records the CME with a linear speed of 848~km s$^{-1}$ in the LASCO FOV, exhibiting a distinct three-component configuration shown in Figure~\ref{fig:WLimages} (f). The CME, which is unrelated to a prominence, originates from the eruption of a HC in our case. There is neither prominence during the eruption nor filament in the source region a few days after the eruption in AIA 304~\AA images. The HC is clearly observed by AIA high-temperature passbands, i.e., 131~\AA\ (with a peak response temperature of $\sim$11~MK) and 94~\AA\ ($\sim$7~MK). An M-class flare is accompanied by the CME, with the 1-8~\AA\ SXR flux curve from the Geostationary Operational Environmental Satellite (GOES) recording the start and end times of the flare at 18:21 and 18:46 UT, respectively.

\begin{figure}
    \centering
    \includegraphics[width=0.8\linewidth]{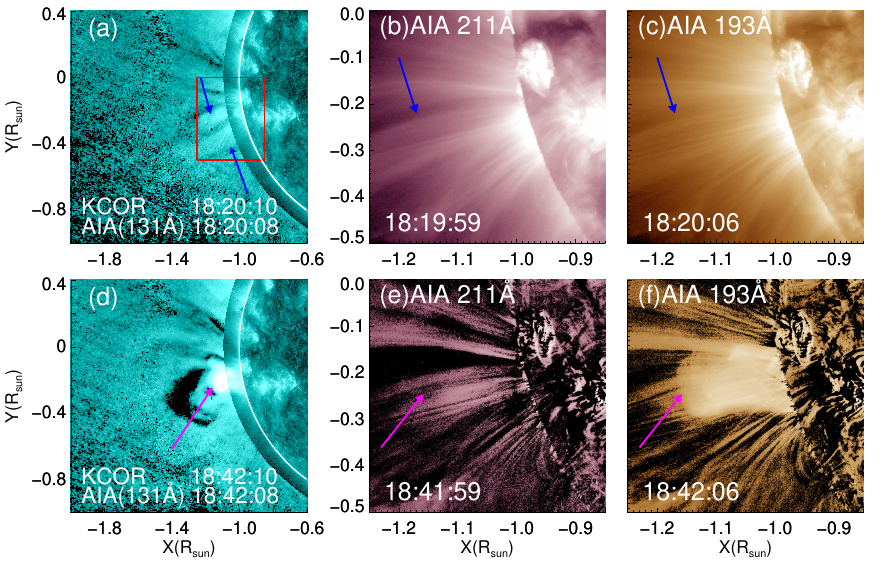}
    \caption{(a) Base-difference K-Cor image at $\sim$18:20~UT, with inset showing the most simultaneous AIA 131 Å base-difference image. The blue arrows mark the pre-eruption bright features, one of them is over-plotted on panels (b) and (c), denoting the background structures in AIA 211~\AA\ and 193~\AA. (b)-(c) Region of interest (marked by red rectangle in panel (a)) of AIA 211~\AA\ and 193~\AA\ direct images at $\sim$18:20~UT. (d) Same as panel (a) but for the exposure at $\sim$18:42~UT, the magenta arrow denotes the HC apex. (e)-(f) The background-subtracted base-difference images of 211~\AA\ and 193~\AA, respectively. The magenta arrows denote the HC apex as observed by AIA, which is overplotted on panel (d).}
    \label{fig:bkgstruct}
\end{figure}

The base-difference K-Cor images, which are obtained by subtracting a pre-event frame at 17:25~UT from the current images, are shown in Figures~\ref{fig:WLimages} (a)-(e). The 131~\AA\ direct images at the nearest time are inlaid in the K-Cor images to show the spatial relationship between the HC and CME structures. We note that there are bright structures appear in the WL observations before the eruption of the HC, as denoted by the blue arrows in Figure~\ref{fig:bkgstruct} (a). They are identified as pre-eruption bright features, which are denser compared to the pre-event corona at 17:25~UT, and become more distinguishable in base-difference images. Figures~\ref{fig:WLimages} (a)-(c) illustrate that the pre-eruption bright features persist until they are completely overlapped by the HC, i.e., the bright core of the CME. We attribute the formation of the pre-eruption features to background corona structures that extend upward, as detected by AIA 211~\AA\ (temperature response peaks at $\sim$2.0~MK) and 193~\AA\ ($\sim$1.6~MK). One of the blue arrows in Figure~\ref{fig:bkgstruct} (a) is also overplotted on panels (b) and (c) to indicate the counterparts of the pre-eruption features in 211~\AA\ and 193~\AA.

In order to distinguish the pre-eruption bright features and the CME bright core, we utilize a set of observations from 211~\AA, 193~\AA, and K-Cor at $\sim$18:42~UT as an illustrative example, as shown in Figures~\ref{fig:bkgstruct} (d)-(f). The magenta arrows mark the location of the HC apex, coinciding with the apex of the CME bright core in panel (d). The residual pre-eruption bright features are still distinguishable in Figures~\ref{fig:bkgstruct} (d)-(f). It clearly shows that the HC enters the K-Cor FOV from the base of the pre-eruption bright features, whose morphologies undergo slight changes throughout the HC eruption. As the HC rises continuously upward, a CME that exhibits a bright core, a dark cavity, and a faint bright front forms in K-Cor images. 

\begin{figure}
    \centering
    \includegraphics[width=0.8\linewidth]{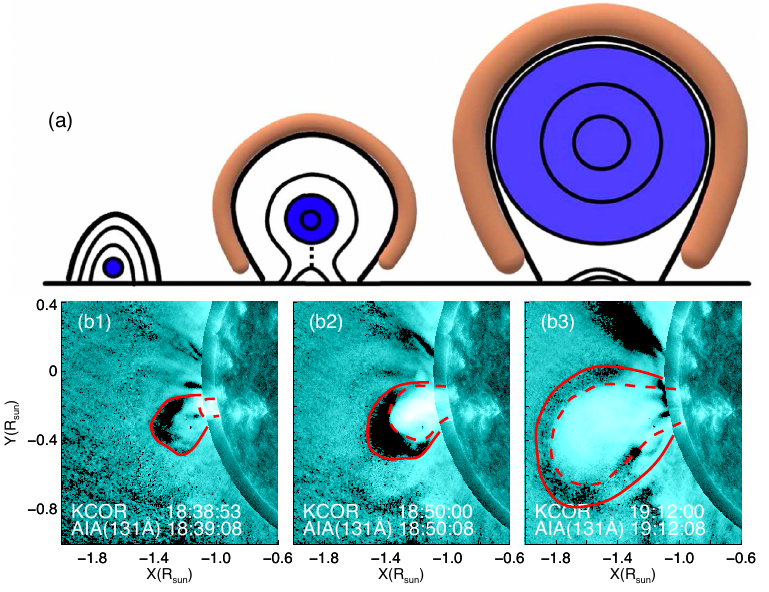}
    \caption{(a) A two-dimensional schematic sketch for the possible new scenario of the intrinsic structure of CMEs \citep[in the absence of filaments, see][Figure 3]{Song_2023ApJ}. (left) The configuration prior to eruption, the circles filled with blue represent the MFR, the thick line represents the outermost coronal loops with potential field, the thin lines represent the sheared magnetic field of the low-density zone (i.e., the cavity). (middle) The surrounding plasma are piled up along the overlying coronal loops as the MFR erupts, forming the CME front (brown). (right) The MFR expands and occupies the low-density zone, resulting in the disappearance of the dark cavity. (b1)-(b3) Base-difference K-Cor images at three time points, showing the evolution of the core and cavity of the CME, corresponding to different stages shown in the upper sketch. The red solid lines indicate the edge of the cavity, and the red dashed lines outline the core.}
    \label{fig:three_part_scenario}
\end{figure}

Additionally, the apparent width of the dark cavity along the HC propagation direction decreases continuously during the eruption, as shown in Figures~\ref{fig:three_part_scenario}, which might indicate the compression of the cavity due to the HC expansion. The observational characteristics of the core and cavity of the CME are somehow consistent with the new explanation of the intrinsic structure of CMEs \citep{Song_2023ApJ}. The schematic sketch in Figure~\ref{fig:three_part_scenario} (a) illustrates the evolution of the three structures of CMEs in the absence of filaments. When an MFR (represented by blue-filled circles) is in equilibrium prior to eruption or slow-rise phase, there is a low-density zone with sheared magnetic field (denoted by thin lines). During the eruption of the MFR, the surrounding plasmas are pile up along the outermost coronal loops (thick line), forming the CME front as denoted in brown color. As the MFR grows and expands during the eruption, it can occupy the low-density zone eventually, leading to the disappearance of the dark cavity. The K-Cor base-difference images in Figures~\ref{fig:three_part_scenario} (b1)-(b3) show clearly that the cavity is occupied gradually by the CME core during the evolution. The cavity and the core are denoted by the red solid lines and dashed lines, respectively.

\begin{figure}[ht!]
    \centering
    \includegraphics[width=0.8\linewidth]{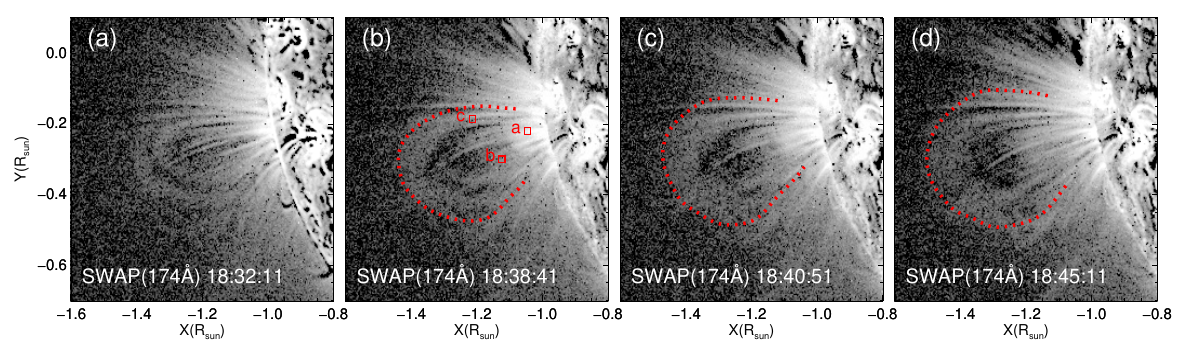}
    \caption{SWAP 174~\AA\ base-difference images, showing the loop-like structures in front of the HC during the eruption. The dotted lines in (b)-(d) denote the outer edge of the loop structures. The red boxes (a, b and c) show the selected sub-regions used for deriving the DEM.}
    \label{fig:SWAP}
\end{figure}

More interestingly, we find that the majority of the cavity region in WL corresponds to bright loop-like structures in SWAP 174~\AA\ at $\sim$1~MK during the evolution. Figure~\ref{fig:SWAP} displays the base-difference images of the SWAP, the red dotted lines in panels (b)-(d) outline the outer edge of the bright loops. The outline of the loop structure is overplotted on the most simultaneous K-Cor images with red pluses, as shown in Figures~\ref{fig:WLimages} (a)-(c). It comes out that the outer edge of the bright loop-like front corresponds well to the location of the outer edge of CME dark cavity. This means that most of the cavity area in K-Cor is filled with the bright loop-like structure observed by SWAP, thanks to the overlapping and larger FOV of SWAP and K-Cor. As a comparison, the cavities enshrouding the prominences are found to be intricate twisted magnetic structures \citep{Habbal_2010ApJ,Habbal_2014ApJ}. However, the bright loop-like structure appears in SWAP 174~\AA\ is possibly less twisted than the HC, even though they are observed from the same perspective. It suggests that the cavity in our event is unlikely to be part of the MFR, differing from that in the prominence-cavity system. 

\subsection{Thermodynamics}\label{subsec:DEM}
\begin{figure}[ht!]
    \centering
    \includegraphics[width=0.9\linewidth]{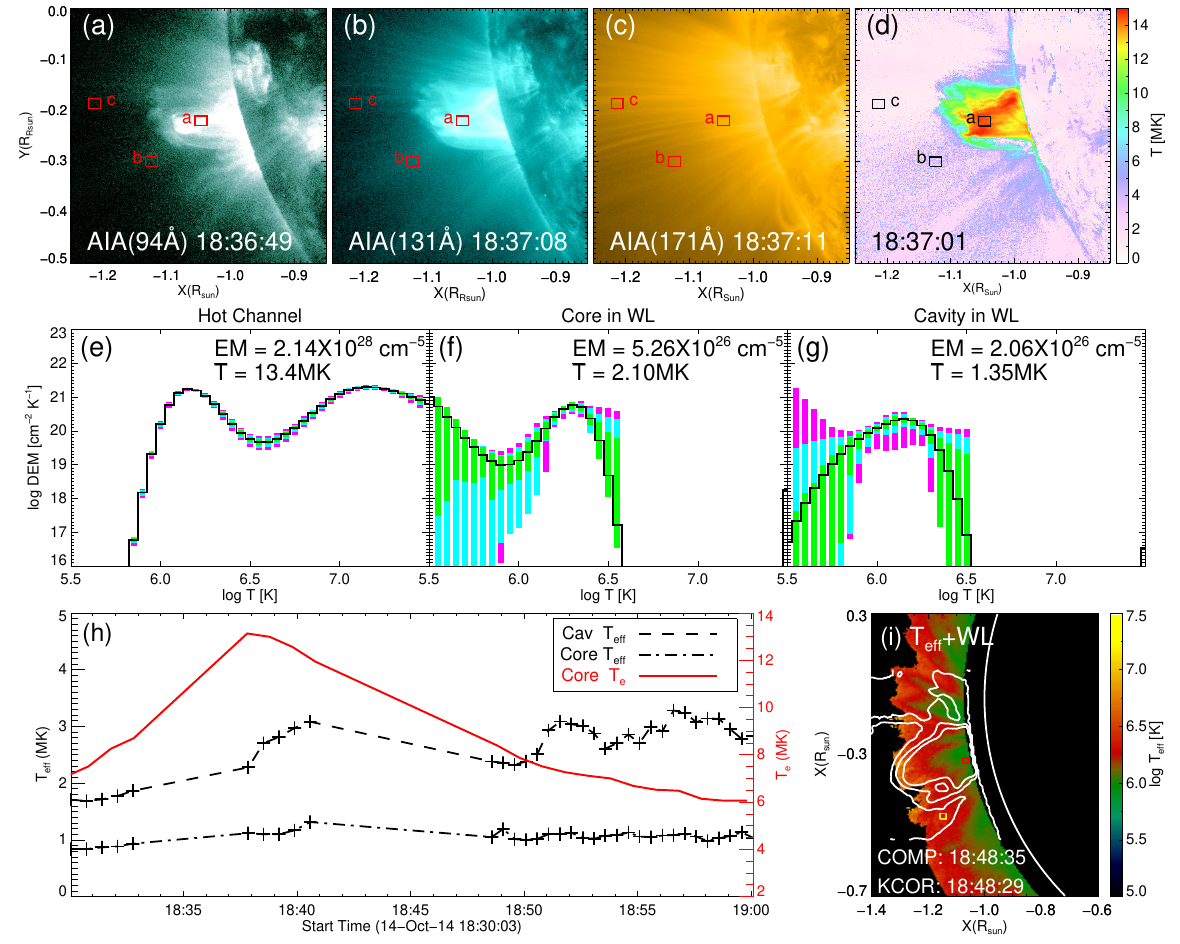}
    \caption{(a)-(c) The direct images of AIA 94~\AA, 131~\AA, and 171~\AA. The sub-regions for DEM analysis corresponding to the HC, pre-eruption bright features, and cavity, are denoted by red boxes with label a, b, and c, respectively. The red boxes are also overlaid in Figure~\ref{fig:WLimages} (a) and Figure~\ref{fig:SWAP} (b). (d) The DEM-weighted temperature map at 18:37~UT. (e)-(f) DEM curves for the three sub-regions. The black solid lines are the best-fitted DEM results. The green rectangles contain 50\% of the MC solutions. The regions including cyan and green rectangles represent 80\% of MC solutions. All the colored rectangles cover 95\% of the MC solutions. (h) Temperatures evolution with time. The black dashed and dash-dot lines (with pluses) show the average effective temperature of the CME core (red box) and cavity (yellow box) sub-regions, marked in panel (i). The red solid curve represents the DEM-weighted temperature within the same sub-region of the CME core. (i) The effective temperature distribution at 18:48~UT. The white contours of K-Cor WL intensity outline the CME core and cavity.}
    \label{fig:Temperature}
\end{figure}

Figures~\ref{fig:Temperature} (a)-(c) display the CME as observed by AIA in different EUV wavelengths at $\sim$18:38~UT, the HC exhibits distinct morphological features owing to its multi-thermal property. Using the DEM method, we have derived the thermal property of the CME from EUV images. The DEM-weighted temperature map at 18:37~UT in Figure~\ref{fig:Temperature} (d) is taken as an example of the results. Furthermore, we selected three distinct sub-regions to illustrate the temperature results. The three sub-regions are outlined by three red boxes in Figures~\ref{fig:Temperature} (a)-(d), corresponding to the HC (box a), the pre-eruption bright feature (box b), and the cavity (box c), respectively. The DEM results of the three sub-regions are shown in Figures~\ref{fig:Temperature} (e)-(g). The black solid curves denote the best-fit DEM solutions to the observed intensity. The colored rectangles indicate the uncertainties in the best-fit solution, as estimated with 100 MC solutions. The green rectangle regions in Figures~\ref{fig:Temperature} (e)-(g) contain 50\% of the MC solutions. The regions including both cyan and green rectangles contain 80\% of the MC solutions. The regions consisting of all the colored rectangles cover 95\% of 100 MC solutions. 

The DEM-weighted temperatures of the three sub-regions are $\sim$13.4~MK, $\sim$2.1~MK, and $\sim$1.3~MK, respectively. Namely, the lowest temperature is located at the CME cavity, the DEM-weighted temperature of the HC is approximately ten times that of the cavity. We note that the DEM solutions for the pre-eruption bright features and the cavity contain great uncertainties due to the poor signal-to-noise near the edge of AIA FOV, implying that we should take the DEM-weighted temperatures of the two structures with caution. Nevertheless, the sub-region of the cavity is located at the leg of the bright EUV loops in SWAP, as shown by the red box c in Figure~\ref{fig:SWAP} (b). The derived temperature of the cavity is somehow consistent with the peak temperature of 174~\AA\ at $\sim$1~MK. It suggests that the DEM result is still reasonable, though with a great uncertainty. The temperature feature coincides with \citet{Bemporad_2007}, in which they reported an increase of the temperature from CME front towards the core as well. The highest temperature of the HC (8.29-10.89~MK), which is regarded as the proxy of MFR, is reported by \cite{Cheng_2012}. It is noteworthy that the lower temperature ($\sim$1~MK) of the HC-associated CME cavity is different from that of the prominence-associated cavity, the latter of which existed prior to the eruption. \citet{Habbal_2010ApJ} reported that the temperature could be 2~MK or even higher within the cavity enshrouding a prominence. \citet{Baksteslicka_2019SoPh} also found that the cavity surrounding the prominence has a relatively high temperature (ranging from 1.67~MK to 2.15~MK) with respect to the lower temperature of the prominence.

The line width observations at \fexiii\ 1074.7/1098.9~nm from CoMP enable us to estimate the effective temperature. We utilize the CoMP data to track the temperature evolution of the CME structures, particularly the cavity whose DEM-weighted temperature has a great uncertainty. Assuming that the term $\xi$ in Equation~\eqref{eq:Teff} equals to zero, the spectral line broadening would only be caused by thermal motion. To this end, the $T_{i,\rm{eff}}$ resulting from Equation~\eqref{eq:Teff} should be considered as the upper limit to the temperature. We take the estimated effective temperature distribution at 18:48~UT as an example to illustrate the result, as shown in Figure~\ref{fig:Temperature} (i). The white contours outline the positions of CME core and cavity observed in K-Cor. Figure~\ref{fig:Temperature} (h) shows the effective temperature evolution of sub-regions in the CME core and cavity, which are respectively marked by the red and yellow box in Figure~\ref{fig:Temperature} (i). The results show that the effective temperature of the cavity is higher than that of the core during the whole evolution.

An interesting result is the evolutionary trends of the effective temperatures. The effective temperature of the cavity generally increases from 1.7~MK to 2.7~MK, while that of the core remains almost constant at $\sim$1~MK during the eruption. We speculate that the most probable reason causing the increased effective temperature of the cavity is the compression due to the expansion of the core. The evidence of the compression has been pointed out in Section~\ref{subsec:morphology}, and will be discussed in detail in Section~\ref{subsec:Kinematics}. For reference, we also derive the electron temperature of the same sub-region in the core from the DEM method, as shown in Figure~\ref{fig:Temperature} (h) by the red line. However, the comparison of the cavity temperatures is not feasible due to the limits of the AIA FOV. Our results show that after the core enters into the red box in Figure~\ref{fig:Temperature} (i) ($\sim$18:30~UT), the DEM-weighted temperature changes greatly in the range of 6~MK to 13~MK, which is much higher than the effective temperature of the CME core derived from the \fexiii\ line that formed at 1.6~MK. It suggests that the heating efficiency for the electrons might be higher than that for the heavy \fexiii\ ions inside the CME core. The different heating efficiencies for different particles due to different heating mechanisms have been demonstrated in some previous works \citep{Cranmer_2020ApJ,Abbo_2016SSRv}.

Furthermore, the difference of the temperature of CME core and cavity might indicate that the magnetic connectivity between the CME core and cavity is weak. If we assume a MFR forms the CME core and cavity simultaneously \citep[e.g.,][]{Howard_2017ApJ}, the length of the magnetic field line connecting the core and cavity could be approximated by the distance ($L_0$) between the cavity outer edge and the center of the core. According to the energy balance equation that considers only the effects of thermal conduction \citep{Gilbert_2015ASSL}, i.e. $3nk\frac{\partial T}{\partial t}=\kappa T^{5/2} \nabla^2T$, where $k$ is the Boltzmann constant, $n$ is the number density, $T$ is the temperature, the conduction coefficient $\kappa$ is a constant. It is in a form of the diffusion equation, then the timescale of the diffusion could be given by $\tau\approx L_0^2/D$, where $D=\kappa T^{5/2}/3nk$. It indicates that the timescale of the thermal conduction along the magnetic field line is about 2.5 minutes. The estimated timescale is much shorter than the timescale of the early evolution stage ($\sim$ 30~minutes). In other words, the temperature of the CME core and cavity shows no significant thermal conduction throughout the eruption, further indicating that the cavity has a different magnetic system from that of the HC.

\citet{Sheoran_2023FrASS} tracked the DEM-weighted and effective temperatures along a slice of the core of another CME, which is associated with neither a prominence nor a HC. Both the DEM-weighted temperature (1.90-2.29~MK) and the effective temperature (0.93-3.71~MK) remain almost constant as the CME evolves, indicating that the expansion of the CME core behaves more like an isothermal than an adiabatic process. Our evolutionary trends of the DEM-weighted temperature and the effective temperature are significantly different from the results in \citet{Sheoran_2023FrASS}. The DEM-weighted temperature of the CME core in their work shows a much narrower temperature distribution peaking at $\sim$1.9~MK, whereas our CME core exhibits emissions over a broader temperature range (Figure~\ref{fig:Temperature} (e)). Technically, the sub-region we selected represents only a minor portion of the HC, as different portions of the HC pass through the red box, the temperatures subsequently vary during evolution. In morphology, the core of our CME event demonstrates significantly greater expansion compared to the event in their work. Therefore, the difference in expansion and the possible difference in heating may lead to more substantial temperature variations in the CME core in the current study, compare to those in their work.


\subsection{Kinematics}\label{subsec:Kinematics}
\begin{figure}[ht!]
    \centering
    \includegraphics[width=0.8\linewidth]{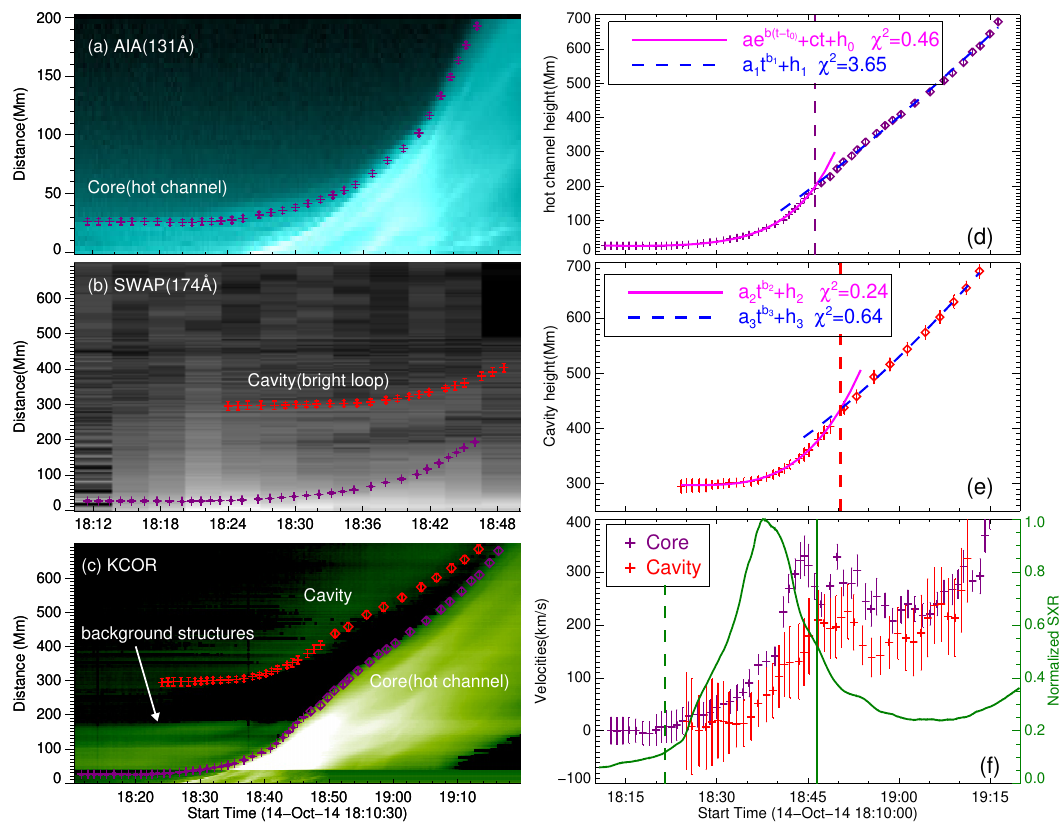}
    \caption{(a) Stack plot of AIA 131~\AA\ data along the slice shown in Figure~\ref{fig:WLimages} (c). The purple pluses denote the measured height of the HC. The height of the HC is overlaid on panels (b) and (c) with the same symbols. (b) The same as panel (a) but for SWAP data, with red pluses indicating the measured height of the top edge of the bright loop structure. (c) The stack plot of the K-Cor data, the measured height of HC and loop edge in panels (a) and (b) are overlaid, with the diamonds showing the extended height profile of CME core and the outer edge of the cavity measured from K-Cor data. The white arrow denotes the pre-eruption bright features as shown in Figure~\ref{fig:bkgstruct} (a). (d)-(e) Fitting of the measured height-time profiles for the CME core and cavity outer edge, respectively, with vertical bars showing the uncertainties in measurements. The magenta and blue lines show the fitting results of different functions for different rising phases, as shown in the top left corner. The purple vertical dashed line indicates the break point between the main-acceleration and the propagation phase of the bright core, the red vertical dashed line denotes that of the cavity. (f) The temporal evolution of the velocity, with vertical bars denoting the uncertainties. The purple pluses represent the velocities of the core, the red ones are for the cavity. The dark green curve is the SXR flux curve, the dashed and solid green vertical lines indicate the start and end times of the associated flare, respectively.}
    \label{fig:slice}
\end{figure}

Given that the effective temperature increase within the narrowing cavity is attributed to compression by the HC expansion, it is essential to investigate the compression process in detail. Figures~\ref{fig:slice} (a)-(c) display the stack plots of AIA, SWAP, and K-Cor, respectively. We visually measure the height-time profile of the HC in AIA FOV, as denoted by the purple pluses in Figure~\ref{fig:slice} (a) with vertical bars indicating the uncertainties. The height-time profile of the bright loop-like structure is measured from SWAP stack plot, as shown in Figure~\ref{fig:slice} (b) with the red pluses. Then the measured height-time profiles of the HC and the loop-like structure are superimposed on the stack plot of K-Cor. Since the K-Cor has a larger FOV than AIA and SWAP, we can extend our track of the eruption outside the AIA or SWAP FOV. The diamonds indicate the extended height-time profile of the CME core (purple diamonds) and that of the cavity outer edge (red diamonds) measured with K-Cor data. It demonstrates that the apex of the EUV loops observed by SWAP at 174~\AA\ coincides with the outer edge of the cavity observed by K-Cor in WL, indicating the density depression of the EUV loops. In addition, the HC starts to catch up with the outer edge of the cavity from behind and becomes the bright CME core in WL. This is evidenced by appearing as the diminishing height difference between the core apex and the cavity outer edge. The pre-existing background structures, the bright CME core and the shrinkage of the cavity can be clearly identified in the K-Cor stack plot along the propagation direction shown in Figure~\ref{fig:slice} (c). We note that the background structure, which is marked by the white arrow in panel (c), remain relatively stable throughout the HC eruption. Here we do not track the evolution of the bright front as it is too faint to be accurately measured in K-Cor FOV. The poor signal will induce a large uncertainty of the measurement. 

The fitting results for the height-time profiles of the CME core and cavity are shown in Figures~\ref{fig:slice} (d) and (e). The symbols indicate the measured height-time profiles, the solid magenta and dashed blue lines are the fitting curves with different functions for different phases. We find that in the AIA FOV, the evolution of the HC exhibits a two-phase behavior (a slow-rise phase and a fast-rise phase), and can be approximated by the nonlinear function with a superposition of the exponential and linear functions (solid magenta line in Figure~\ref{fig:slice} (d)). However, when we extend the fitting into higher altitude, the nonlinear function shows a faster rise in the K-Cor FOV than the measured height-time profile. We therefore use a power law function with a index of 1.32 to yield a better fit for the extended part of the height-time profile of the core. Note that the variable $t$ we used for all the functions represents the time difference between the measured points and the first point. The break point, where the exponential function begins to exceed the measured height, marks the end time of the main-acceleration phase. The break point between the two functions is delineated by the purple vertical line in Figure~\ref{fig:slice} (d). The fitting for the cavity evolution yields similar results to those of the HC, i.e., it can only be best fitted individually by two functions for different phases, with a break point indicating the start of the last phase. However, the functions are different from those used for fitting the height-time profile of the HC, the best-fit functions for the cavity are both power law, though with different power-law indices. The fitting function for the height-time profile of the cavity measured from SWAP stack plot has a power-law index of 3.4, while the corresponding index for the extended portion of the profile measured from K-Cor stack plot is 1.5. The break point between the two functions is indicated by the red vertical line in Figure~\ref{fig:slice} (e), which is approximately 4 minutes later than that of the core.

The speed evolution of the HC (purple pluses with vertical bars showing the uncertainty) and the cavity (red pluses with vertical bars), as well as the SXR flux curve of the flare (green curve) are shown in Figure~\ref{fig:slice} (f). We find that the three-phase evolution of the HC is temporally closely related to the SXR flux of the associated flare, demonstrating that these two phenomenons are correlated with each other by the same physical mechanism, e.g., the conversion of magnetic energy to kinetic and thermal energy via magnetic reconnection \citep{Cheng_2013ApJ}. The green vertical dashed and solid lines indicate the recorded start and end times of the associated flare, respectively. Before the start of the impulsive phase of the flare, the speed of the HC increases slowly to $\sim$10~km s$^{-1}$ within 10 minutes. As the impulsive phase of the flare starts, the speed of the HC increases rapidly from $\sim$10~km s$^{-1}$ to $\sim$330~km s$^{-1}$ in 25 minutes. Finally, the HC evolves with a residual acceleration in the decay phase of the flare. As for the cavity, we find that the speed evolution is similar to but always lags behind the HC. Moreover, the speed of the cavity (peaks at 220~km s$^{-1}$) is always slower than that of the HC. Thus, similar to the conclusion in \cite{Cheng_2013ApJ}, we argue that the HC acts as the main driver of the CME formation and eruption. The cavity is pushed and compressed by the eruption of the HC, and consequently, heating the plasma inside the cavity. We note that the linear speed of the CME in the LASCO FOV provided by CDAW is 848~km s$^{-1}$, which is much higher than the maximum speed we measured in the inner corona. It means the CME may propagate continuously with a residual acceleration. The appearance of the bright front in LASCO can be attributed to the accumulation of the plasma from the inner corona to the outer corona. The bright front in K-Cor is less distinct than in LASCO, possibly because the low speed of the CME in the inner corona prevents it from accumulating sufficient plasma to form a pronounced bright front.

\begin{figure}[ht!]
    \centering
    \includegraphics[width=0.9\linewidth]{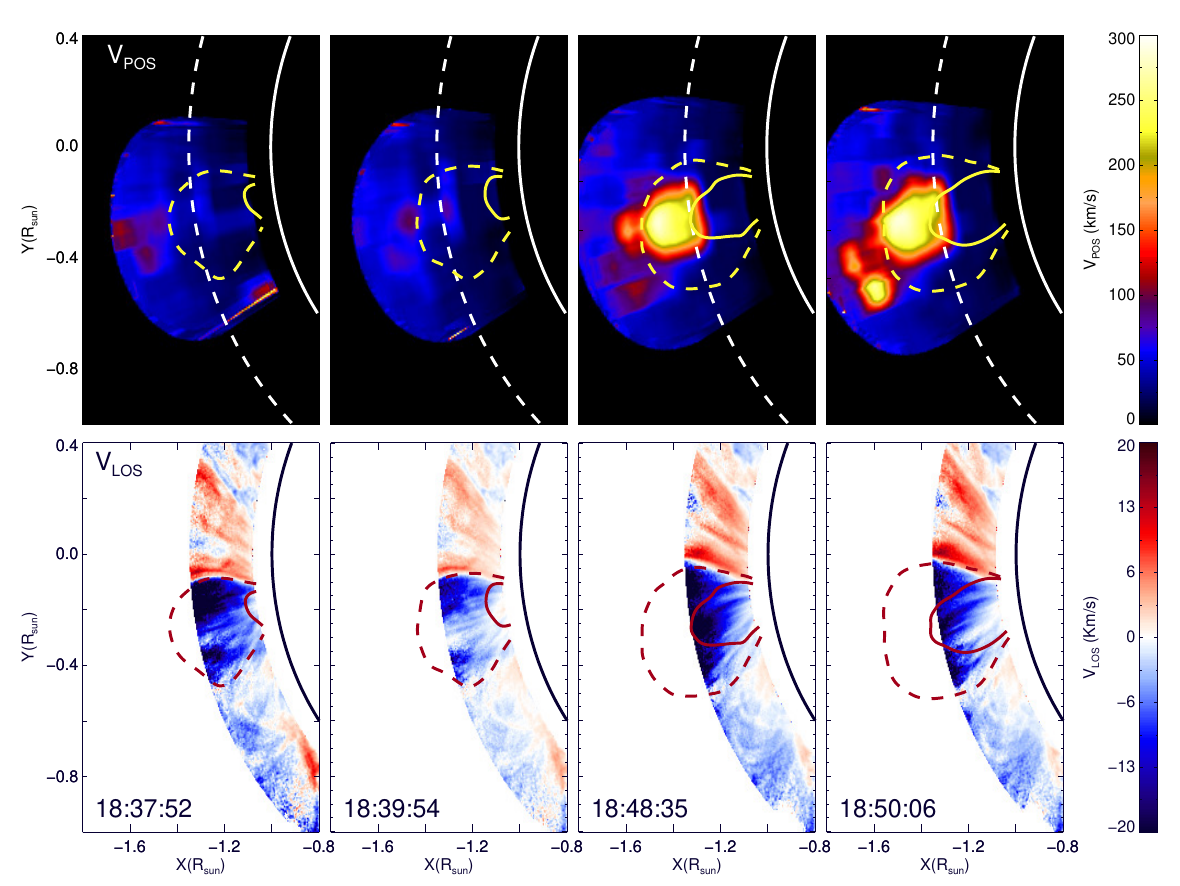}
    \caption{Upper row: Four snapshots of the projected radial velocity in the POS derived from CCM method. The yellow solid and dashed lines delineate the CME core and the cavity outer edge respectively observed by K-Cor. Bottom: the CoMP Doppler velocity observations along the LOS during the eruption. The core and cavity are denoted by the red lines. The white solid lines indicate the solar limb and the dashed lines represent the outer FOV of the CoMP.}
    \label{fig:3d_speed}
\end{figure}

By using the CCM, we obtained the average radial speed distribution in the POS, as shown in the top panels of Figure~\ref{fig:3d_speed}. The white dashed lines mark the outer FOV of the COMP, and the white solid lines mark the solar limb. The yellow solid and dashed lines denote the core and the cavity edge observed in WL, respectively. The results indicate that the highest speed of the CME along the radial direction is located at the core apex during the impulsive acceleration phase of the HC, with the maximum velocity exceeding 300~km s$^{-1}$. The value is consistent with the velocity derived from the height-time profile of the CME core. Moreover, the radial speed distribution can demonstrate again that the HC drives the CME eruption and compresses the cavity. In addition to the 2D radial speed distribution in the POS, the observations from CoMP can provide us the information of the Doppler velocity along the LOS. The Doppler velocitiy observations at different times are shown in the bottom panels of Figure~\ref{fig:3d_speed}, with red solid line and dashed line indicating the core and cavity edge in WL, respectively. We note that the Doppler shift in the cavity is dominated by blue shift with a maximum speed of about 20~km s$^{-1}$, whereas the core is much less blue shifted. It is reasonable because the leg of the HC is found to be fixed mostly in the photosphere, only the top portion of the HC stretches upward. The Doppler velocity implies that the HC, i.e., the CME bright core, may propagate outward around POS along with expansion.

\section{Summary and Discussion}\label{sec:Summary}
Thanks to the overlapping and large FOV of SWAP and K-Cor, and multi-wavelength observations together with AIA and CoMP, we have investigated the early evolution of a HC-associated CME eruption on 2014 October 14. In particular, we have revealed the morphological, thermodynamic, and kinematic properties of the CME dark cavity in the inner corona. To our knowledge, it is the first time that we have found that the WL cavity is mainly filled by the observed cool loop structures in 174~\AA, as well as the observational evidence of the heating inside the cavity due to CME core compression in the early evolution stage. Additionally, we have demonstrated the distinct properties between cavity associated with prominence and that related to the HC.

The HC is observed erupting from the solar limb by AIA high-temperature passbands (131~\AA, 94~\AA). As it rises upward continuously, the HC evolves from the base of the pre-existing bright features, which correspond to denser background corona structures that are well imaged in AIA 211~\AA\ and 193~\AA. After the HC enters K-Cor FOV, it appears as the bright CME core. The DEM-weighted temperature decreases from the core to the cavity, indicating the multi-thermal component of the CME. The highest temperature of $\sim$13.4~MK in the CME core, i.e., the HC, is approximately ten times that of the cavity ($\sim$1.3~MK). On the other hand, we note that the cavity is continuously compressed by the expanding CME core, The effective temperature of \fexiii\ in the cavity derived from the line width measurements generally increases with time, which can be considered as a strong evidence of the heating in the cavity via compression. Furthermore, the high speed region at the core apex along the HC main propagation direction is another evidence of cavity compression by the expansion of the HC.

In the traditional scenario of the CME three components, the whole cavity is regarded as the MFR, the CME bright core corresponds to the dense filament plasma supported by the MFR against the gravity, and the bright leading front is due to the accumulation of the plasma along the MFR boundary. The cavity volume should increase with the expansion of the MFR as predicted by the traditional scenario. On the contrary, our analyses reveal that the width of the dark cavity decreases due to the expansion of the HC. To this end, the evolutionary characteristics of the present event support the new explanation of the CME cavity -- it may not be a part of the MFR structure, but presumably corona loops with lower density overlying the MFR. The formation of the low-density zone between the MFRs and the front has been predicted by the model in \cite{Haw_2018ApJ}. \cite{Song_2025ApJ} has given some observation evidence implying the nature of the cavity to be a low-density zone between the CME front and MFR. Regarding the magnetic structure of the cavity, it should be understood with further magnetic measurements and analysis methods.

Another interesting finding concerns the distinct physical properties between the cavity associated with a prominence and that associated with a HC. Magnetically, the cool bright loop-like structures in SWAP 174~\AA\ are less complex and twisted than HCs or the turbulent-like structures typically observed in pre-existing prominence cavities. On the contrary, the magnetohydrodynamic simulation presented by \cite{Fan_2019FrASS} indicated that a bright loop-like structure in 171~\AA\ corresponds to the CME leading edge for a traditional prominence-cavity system, while the cavity is threaded by the twisted field lines carrying prominence dips. Nevertheless, \cite{Shaik_2024ApJ} demonstrates that an erupting prominence-cavity outlines the projection of the three-dimensional structure of the MFR. \cite{Sarkar_2019ApJ} illustrates the cavity enshrouding a prominence evolves into the cavity in WL. Additionally, the flows in the POS or along the LOS trace out the magnetic field lines. As reported by \citet{Gibson_2015ASSL}, it is common to observe flows along the LOS within prominence cavities, and occasionally the counter-streaming in the form of nested ring-like structures when the axis of the MFR orients along the LOS \citep{Baksteslicka_2013ApJ}. The POS flows along prominence-cavity interface horns indicate the magnetic field geometry of the prominence cavity \citep{Schimit_2013ApJ}. Our observations reveal no pronounced horn-like POS flows, which indicates the magnetic structure of the cavity is probably different from that of the prominence cavity. 

In terms of temperature, the DEM-weighted temperature of the cavity surrounding the HC is $\sim$1.3~MK. This is lower than the temperature of the cavity ($\ge$2~MK) surrounding the prominence, either derived from AIA observations with DEM method \citep{Baksteslicka_2019SoPh} or spectral imaging analysis \citep{Habbal_2010ApJ}. \cite{Bazin_2013SoPh} combined SWAP images and simultaneous eclipse spectral observations to study the prominence cavity regions, they tend to confirm that prominence cavities are filled with hot plasma at approximately 2~MK or even higher. Furthermore, these two cavity systems have opposite core-cavity temperature relationships -- the HC is hotter than the surrounding cavity, while the prominence is cooler than its cavity.

An associated M-class flare is occurred with the eruption of the CME, according to the GOES SXR flux curve. The pre-flare phase, the impulsive phase and the decay phase of the flare are temporally aligned with the evolution of the CME core. It undergoes a slow-rise phase, a main-acceleration phase, and a simple propagation phase which has a small residual acceleration. The correlation of the CME eruption and the flare SXR flux has been reported in many previous works \citep{Cheng_2013ApJ,Zhang_2004,Cheng_2010}. To date, it is widely accepted that the CMEs are coupled together with the associated flare by the same physical mechanism, especially during the main energy release phase, presumably via magnetic reconnection \citep{Lin_2000JGR}. The slow-rise phase is suggested to be caused by the slow magnetic reconnection, while the torus instability or the magnetic reconnection producing flares can lead to the onset of the impulsive acceleration of the HC \citep{Green_2018SSRv}. However, the trigger of the main-acceleration of CME is still an open issue, and is not the main focus of the present work. On the other hand, the velocity of the HC is always faster than that of the cavity. We conclude that the eruption of the HC drives the formation and acceleration of the CME, at least in the early evolution stage. The cavity is pushed and compressed by the expansion of HC, which is supported by both morphological and dynamical analyses of the CME. \cite{Cheng_postimpulsive_2010ApJ} suggested that the positive residual acceleration may continue to exist after the impulsive acceleration phase. It also occurred in the evolution of the present CME, the maximal velocity of the HC in the inner corona is about 330~km s$^{-1}$. However, the linear speed of 848~km s$^{-1}$ is recorded by CDAW in LASCO/C2 FOV with a clear CME bright front. We speculate that the formation of the bright front in C2, which is hard to be clearly identified in K-Cor, is attributed to the continuous residual acceleration of the CME and the pileup of the plasma along the cavity border.

\section{Acknowledgments}
The authors acknowledge the use of data from the SDO, PROBA2, MLSO, and SOHO. This work is supported by National Key R\&D Program of China 2022YFF0503003 (2022YFF0503000), the Strategic Priority Research Program of the Chinese Academy of Sciences, Grant No. XDB0560000, NSFC grants (Nos.12233012, 12203102). This work benefits from the discussions of the ISSI-BJ Team ``Solar eruptions: preparing for the next generation multi-waveband coronagraphs".








\bibliography{sample631}{}
\bibliographystyle{aasjournal}

\end{document}